\documentclass[twocolumn,showpacs,preprintnumbers,amsmath,amssymb]{revtex4}

\usepackage{graphicx}
\usepackage{dcolumn}
\usepackage{bm}

\usepackage{amssymb}
\usepackage{amsmath}

\begin{document}


\title{Cosmological scenario based on the first and second laws of thermodynamics: Thermodynamic constraints on a generalized cosmological model}

\author{Nobuyoshi {\sc Komatsu}}  \altaffiliation{E-mail: komatsu@se.kanazawa-u.ac.jp} 
\affiliation{Department of Mechanical Systems Engineering, Kanazawa University, Kakuma-machi, Kanazawa, Ishikawa 920-1192, Japan}

\date{\today}

\begin{abstract}
The first and second laws of thermodynamics should lead to a consistent scenario for discussing the cosmological constant problem.
In the present study, to establish such a thermodynamic scenario, cosmological equations in a flat Friedmann--Lema\^{i}tre--Robertson--Walker universe were derived from the first law, using an arbitrary entropy $S_{H}$ on a cosmological horizon.
Then, the cosmological equations were formulated based on a general formulation that includes two extra driving terms, $f_{\Lambda}(t)$ and $h_{\textrm{B}}(t)$, which are usually used for, e.g., time-varying $\Lambda (t)$ cosmology and bulk viscous cosmology, respectively.
In addition, thermodynamic constraints on the two terms are examined using the second law of thermodynamics, extending a previous analysis [Phys. Rev. D \textbf{99}, 043523 (2019)].
It is found that a deviation $S_{\Delta}$ of $S_{H}$ from the Bekenstein--Hawking entropy plays important roles in the two terms.
The second law should constrain the upper limits of $f_{\Lambda}(t)$ and $h_{\textrm{B}}(t)$ in our late Universe.
The orders of the two terms are likely consistent with the order of the cosmological constant $\Lambda_{\textrm{obs}}$ measured by observations.
In particular, when the deviation $S_{\Delta}$ is close to zero, $h_{\textrm{B}}(t)$ and $f_{\Lambda}(t)$ should reduce to zero and a constant value (consistent with the order of $\Lambda_{\textrm{obs}}$), respectively, as if a consistent and viable scenario could be obtained from thermodynamics.

\end{abstract}


\maketitle

\section{Introduction} 
\label{Introduction}

Lambda cold dark matter ($\Lambda$CDM) models, which assume a cosmological constant $\Lambda$ and dark energy, can explain an accelerated expansion of the late Universe \cite{PERL1998_Riess1998,Planck2018,Hubble2017}.
However, the $\Lambda$CDM model suffers from several theoretical problems \cite{Weinberg1989}.
For example, it is well-known that $\Lambda$ measured by observations is approximately $60 \sim 120$ orders of magnitude smaller than the theoretical value obtained from quantum field theory \cite{Weinberg1989,Pad2003,Barrow2011,Bao2017}.
To resolve those problems, astrophysicists have proposed various cosmological models, 
such as time-varying $\Lambda (t)$ cosmology \cite{FreeseOverduin,Nojiri2006etc,Valent2015Sola2019,Sola_2009-2022}, bulk viscous cosmology \cite{BarrowLima,BrevikNojiri,EPJC2022}, and 
creation of CDM (CCDM) models \cite{Prigogine_1988-1989,Lima1992-1996,LimaOthers2023,Freaza2002Cardenas2020}. 
In addition, thermodynamic scenarios based on the holographic principle \cite{Hooft-Bousso}, such as entropic cosmology \cite{EassonCai,Basilakos1,Koma45,Koma6,Koma78,Koma9,Neto2022,Gohar2024}  
and holographic cosmology \cite{Padmanabhan2004,ShuGong2011,Padma2012AB,Cai2012,Hashemi,Moradpour,Koma10,Koma11,Koma12,Koma18,Pad2017,Tu2018,Tu2019,HDE,Krishna20172019,Mathew2022,Chen2022,Luciano,Mathew2023,Mathew2023b,Koma14,Koma15,Koma16,Koma17,Koma19,Koma20}, 
have been examined extensively \cite{Koma21,Cai2005,Cai2011,Dynamical-T-2007,Dynamical-T-20092014,Sheykhi1,Sheykhi2Karami,Mirza2015,Silva2015,Santos2022,Sheykhia2018,ApparentHorizon2022,Cai2007,Cai2007B,Cai2008,Sanchez2023,Nojiri2024,Odintsov2023ab,Odintsov2024,Mohammadi2023,Odintsov2024B,Nojiri2024B}.

In the thermodynamic scenarios, black hole thermodynamics \cite{Hawking1Bekenstein1} is applied to a cosmological horizon, which is assumed to have an associated entropy and an approximate temperature \cite{EassonCai}.
In those models \cite{Cai2007,Cai2007B,Cai2008,Sanchez2023,Nojiri2024,Odintsov2023ab,Odintsov2024,Mohammadi2023,Odintsov2024B,Nojiri2024B}, cosmological equations are derived from the first law of thermodynamics using a dynamical Kodama--Hayward temperature \cite{Dynamical-T-2007,Dynamical-T-20092014} and various forms of black hole entropy (including Bekenstein--Hawking entropy \cite{Hawking1Bekenstein1}).
For example, modified cosmological equations, which include extra driving terms, can be formulated by applying a power-law-corrected entropy \cite{Das2008,Radicella2010}, Tsallis--Cirto entropy \cite{Tsallis2012}, Tsallis--R\'{e}nyi entropy \cite{Czinner1Czinner2}, Barrow entropy \cite{Barrow2020}, and a generalized six-parameter entropy \cite{Nojiri2022}.
In addition, an arbitrary entropy on the horizon can be used to derive a generalized cosmological model from the first law, as examined by Odintsov \textit{et al.} \cite{Odintsov2024}.
We expect that the second law of thermodynamics constrains driving terms in the generalized model, as if the cosmological constant problem could be discussed from a thermodynamics viewpoint.
(For the first law, see, e.g., the previous works of Akbar and Cai \cite{Cai2007,Cai2007B} and Cai \textit{et al.} \cite{Cai2008} and the recent works of S\'{a}nchez and Quevedo \cite{Sanchez2023}, Nojiri \textit{et al.} \cite{Nojiri2024}, Odintsov \textit{et al.} \cite{Odintsov2023ab,Odintsov2024,Odintsov2024B}, and the present author \cite{Koma21}.
 See also a recent review \cite{Nojiri2024B}.)

In fact, the present author has examined thermodynamic constraints on an extra driving term in holographic equipartition models, similar to a time-varying $\Lambda (t)$ cosmology \cite{Koma11,Koma12}.
However, theoretical backgrounds and cosmological equations for the model are different from those for the generalized cosmological model derived from the first law of thermodynamics.
In addition, the generalized cosmological model should include two different driving terms, such as $f_{\Lambda}(t)$ and $h_{\textrm{B}}(t)$, unlike for the holographic equipartition model.
(Usually, $f_{\Lambda}(t)$ and $h_{\textrm{B}}(t)$ are used for, e.g., time-varying $\Lambda (t)$ cosmology and bulk viscous cosmology, respectively, based on a general formulation \cite{Koma16,Koma21}.)
The two terms should be related to a deviation of horizon entropy from the Bekenstein--Hawking entropy.
We expect that the generalized cosmological model provides a consistent thermodynamic scenario; that is, the generalized model is derived from the first law, whereas the second law constrains the two terms in the model.
The generalized cosmological model and thermodynamic constraints on the two driving terms have not yet been examined from those viewpoints.
(Note that the second law itself has been examined in, e.g., Refs.\ \cite{Odintsov2024,Nojiri2024B}.)

In this context, we examine thermodynamic constraints on the two terms in the generalized cosmological model, extending previous work \cite{Koma11,Koma12}.
In the present study, cosmological equations are derived from the first law of thermodynamics using an arbitrary entropy on the cosmological horizon, in accordance with Ref.\ \cite{Odintsov2024}.
The original cosmological equations implicitly include extra driving terms. 
Therefore, the cosmological equations are systematically formulated again, based on a general formulation that explicitly includes two extra driving terms, $f_{\Lambda}(t)$ and $h_{\textrm{B}}(t)$.
In addition, we universally examine the thermodynamic constraints on the two driving terms using the second law of thermodynamics.
The present study should contribute to a better understanding of thermodynamic scenarios and may provide new insights into the discussion of the cosmological constant problem.
Inflation of the early universe is not discussed here because we focus on the late universe.

The remainder of the present article is organized as follows.
In Sec.\ \ref{Cosmological equations}, a general formulation for cosmological equations is reviewed.
In Sec.\ \ref{Entropy and temperature}, an associated entropy and an approximate temperature on a cosmological horizon are introduced.
In Sec.\ \ref{The first law}, cosmological equations are derived from the first law of thermodynamics using an arbitrary entropy on the horizon.
In addition, based on the general formulation, the cosmological equations are systematically formulated. 
In Sec.\ \ref{Thermodynamic constraints}, thermodynamic constraints on the two terms in the present model are examined based on the second law of thermodynamics.
Finally, in Sec.\ \ref{Conclusions}, the conclusions of this study are presented.

\section{General cosmological equations in a flat FLRW universe} 
\label{Cosmological equations}

The present study considers a flat Friedmann--Lema\^{i}tre--Robertson--Walker (FLRW) universe.
In addition, an expanding universe is assumed from observations \cite{PERL1998_Riess1998,Hubble2017,Planck2018}.
A general formulation for cosmological equations was previously examined in Refs.\ \cite{Koma6,Koma9,Koma14,Koma15,Koma16} and recently examined in Ref.\ \cite{Koma21}.
In this section, we introduce a general formulation using the scale factor $a(t)$ at time $t$, in accordance with those works.

The general Friedmann, acceleration, and continuity equations are written as \cite{Koma21}
\begin{equation}
 H(t)^2      =  \frac{ 8\pi G }{ 3 } \rho (t)    + f_{\Lambda}(t)            ,                                                 
\label{eq:General_FRW01} 
\end{equation} 
\begin{align}
  \frac{ \ddot{a}(t) }{ a(t) }     &= -  \frac{ 4\pi G }{ 3 }  \left ( \rho (t) +  \frac{3 p(t)}{c^2} \right )                   +   f_{\Lambda}(t)    +  h_{\textrm{B}}(t)  , 
\label{eq:General_FRW02}
\end{align}
\begin{equation}
       \dot{\rho} + 3  H \left ( \rho (t) +  \frac{p(t)}{c^2} \right )       =    -  \frac{3}{8 \pi G }   \dot{f}_{\Lambda}(t)      +    \frac{3 }{4 \pi G}     H h_{\textrm{B}}(t)              , 
\label{eq:drho_General}
\end{equation}
with the Hubble parameter $H(t)$ defined as 
\begin{equation}
   H(t) \equiv   \frac{ da/dt }{a(t)} =   \frac{ \dot{a}(t) } {a(t)}  ,
\label{eq:Hubble}
\end{equation}
where $G$, $c$, $\rho(t)$, and $p(t)$ are the gravitational constant, the speed of light, the mass density of cosmological fluids, and the pressure of cosmological fluids, respectively.
Also, $f_{\Lambda}(t)$ and $h_{\textrm{B}}(t)$ are extra driving terms \cite{Koma14,Koma21}.
Usually, $f_{\Lambda}(t)$ is used for a $\Lambda (t)$ model, similar to $\Lambda(t)$CDM models, whereas $h_{\textrm{B}}(t)$ is used for a bulk-viscous-cosmology-like model, similar to bulk viscous models and CCDM models \cite{Koma14,Koma16,Koma21}.
In this paper, the two terms are phenomenologically assumed and are considered to be related to an associated entropy on a cosmological horizon, as examined later.
The general continuity given by Eq.\ (\ref{eq:drho_General}) can be derived from Eqs.\ (\ref{eq:General_FRW01}) and (\ref{eq:General_FRW02}), because only two of the three equations are independent \cite{Ryden1}. 
In addition, subtracting Eq.\ (\ref{eq:General_FRW01}) from Eq.\ (\ref{eq:General_FRW02}) yields \cite{Koma21} 
\begin{equation}
    \dot{H} = - 4\pi G  \left ( \rho +  \frac{p}{c^2} \right )  + h_{\textrm{B}}(t)   .
\label{eq:dotH}
\end{equation}
These equations are used in Sec.\ \ref{The first law} to examine cosmological equations derived from the first law of thermodynamics.

Equation\ (\ref{eq:drho_General}) indicates that the right side of the general continuity equation is usually non-zero, as discussed in Refs.\ \cite{Koma9,Koma20,Koma21}.
However, when both $f_{\Lambda} (t) = \Lambda / 3$ and $ h_{\textrm{B}} (t) =0$ are considered, the continuity equation reduces to the standard continuity equation, namely $\dot{\rho} + 3  H [ \rho +  (p/c^2)] =0$. 
Exactly speaking, Eq.\ (\ref{eq:drho_General}) reduces to the standard continuity equation when the following relation is satisfied:
\begin{equation}
       h_{\textrm{B}}(t)     =   \frac{  \dot{f}_{\Lambda}(t) }{ 2 H}     . 
\label{eq:zero_hB_fL_org}
\end{equation}
In the present study, we derive a generalized cosmological model from the first law of thermodynamics, by applying the standard continuity equation, as examined in previous works \cite{Cai2007,Cai2007B,Cai2008,Sanchez2023,Nojiri2024,Odintsov2023ab,Odintsov2024,Odintsov2024B,Nojiri2024B}.
That is, we assume that Eq.\ (\ref{eq:zero_hB_fL_org}) holds.
We can confirm that substituting Eq.\ (\ref{eq:zero_hB_fL_org}) into Eq.\ (\ref{eq:drho_General}) gives the standard continuity equation, written as
\begin{equation}
\dot{\rho} + 3  H \left ( \rho +  \frac{p}{c^2} \right )       =  0 .
\label{eq:drho_Zero}
\end{equation}
The right side of this equation is zero but includes $f_{\Lambda} (t)$ and $h_{\textrm{B}} (t)$ implicitly.

It should be noted that coupling Eq.\ (\ref{eq:General_FRW01}) with Eq.\ (\ref{eq:General_FRW02}) yields the cosmological equation \cite{Koma14,Koma15,Koma16} given by 
\begin{align}
    \dot{H}    &= - \frac{3}{2} (1+w)  H^{2}  +  \frac{3}{2}   (1+w)  f_{\Lambda}(t)                 + h_{\textrm{B}}(t)   \notag \\
                   &= -  \frac{3}{2} (1+w)  H^{2} \left ( 1- \frac{f_{\Lambda}(t)}{H^{2}} \right )    + h_{\textrm{B}}(t)   ,
\label{eq:Back2}
\end{align}
where $w$ represents the equation of the state parameter for a generic component of matter, which is given as $w = p/(\rho  c^2)$ \cite{Koma21}.
For a $\Lambda$-dominated universe and a matter-dominated universe, the values of $w$ are $-1$ and $0$, respectively.
In this paper, $w > -1$ is considered because $f_{\Lambda}(t)$ can behave as a varying cosmological-constant-like term instead of $w = -1$.
(Note that $w$ is retained for generality.)
For example, when both $f_{\Lambda} (t) = \Lambda / 3$ and $ h_{\textrm{B}} (t) =0$ are considered, the general formulation reduces to that for $\Lambda$CDM models.
The order of the density parameter $\Omega_{\Lambda}$ for $\Lambda$ is $1$, based on the Planck 2018 results \cite{Planck2018}.
Here $\Omega_{\Lambda}$ is defined by $\Lambda /(3 H_{0}^{2})$, and $H_{0}$ is the current Hubble parameter.
Therefore, the order of the cosmological constant term measured by observations, namely $O( \Lambda_{\textrm{obs}}/3)$ should be written as \cite{Koma10,Koma11,Koma12}
 \begin{equation}
  O \left ( \frac{\Lambda_{\textrm{obs}}}{3}  \right )     \approx   O  \left ( H_{0}^{2}  \right ) . 
\label{L_order_0}
\end{equation} 
The orders of two extra driving terms are discussed later.

\section{Entropy and temperature on the cosmological horizon} 
\label{Entropy and temperature}

In thermodynamic scenarios, a cosmological horizon is assumed to have an associated entropy and an approximate temperature \cite{EassonCai}.
In this section, the entropy $S_{H}$ and the temperature $T_{H}$ on the horizon are introduced in accordance with previous works \cite{Koma11,Koma12,Koma17,Koma18,Koma19,Koma20,Koma21}.

First, we review a form of the Bekenstein--Hawking entropy as an associated entropy on the cosmological horizon because it is the most standard approach \cite{Koma19,Koma20,Koma21}.
In the present paper, a flat FLRW universe is considered and, therefore, the Hubble horizon is equivalent to the apparent horizon of the universe.
Based on the form of the Bekenstein--Hawking entropy, the entropy $S_{\rm{BH}}$ is written as \cite{Hawking1Bekenstein1}  
\begin{equation}
S_{\rm{BH}}  = \frac{ k_{B} c^3 }{  \hbar G }  \frac{A_{H}}{4}   ,
\label{eq:SBH}
\end{equation}
where $k_{B}$ and $\hbar$ are the Boltzmann constant and the reduced Planck constant, respectively.
The reduced Planck constant is defined by $\hbar \equiv h/(2 \pi)$, where $h$ is the Planck constant \cite{Koma11,Koma12}.
$A_{H}$ is the surface area of the sphere with a Hubble horizon (radius) $r_{H}$ given by
\begin{equation}
     r_{H} = \frac{c}{H}   .
\label{eq:rH}
\end{equation}
Substituting $A_{H}=4 \pi r_{H}^2 $ into Eq.\ (\ref{eq:SBH}) and applying Eq.\ (\ref{eq:rH}) yields
\begin{equation}
S_{\rm{BH}}  = \frac{ k_{B} c^3 }{  \hbar G }   \frac{A_{H}}{4}       
                  =  \left ( \frac{ \pi k_{B} c^5 }{ \hbar G } \right )  \frac{1}{H^2}  
                  =    \frac{K}{H^2}    , 
\label{eq:SBH2}      
\end{equation}
where $K$ is a positive constant given by
\begin{equation}
  K =  \frac{  \pi  k_{B}  c^5 }{ \hbar G } . 
\label{eq:K-def}
\end{equation}
Differentiating Eq.\ (\ref{eq:SBH2}) with regard to $t$ yields \cite{Koma11,Koma12}
\begin{equation}
\dot{S}_{\rm{BH}}  
                          = \frac{d}{dt}   \left ( \frac{K}{H^{2}} \right )  =  \frac{-2K \dot{H} }{H^{3}}                  .
\label{eq:dSBH}      
\end{equation}
Cosmological observations indicate that $H >0$ and $\dot{H} < 0$ (see, e.g., Ref.\ \cite{Hubble2017}).
Accordingly, in our Universe, $\dot{S}_{\rm{BH}}$ should be positive as follows \cite{Koma11,Koma12}:
\begin{equation}
\dot{S}_{\rm{BH}}   =  \frac{-2K \dot{H} }{H^{3}}  >  0  .  
\label{eq:dSBH_2}      
\end{equation}
Of course, various forms of black hole entropy \cite{Tsallis2012,Czinner1Czinner2,Barrow2020,Nojiri2022} have been proposed and these entropies can be interpreted as an extended version of $S_{\rm{BH}}$.
In this study, we consider an arbitrary form of entropy $S_{H}$ on the Hubble horizon and derive a generalized cosmological model from the first law of thermodynamics, as examined in Sec.\ \ref{The first law}.

Next, we introduce an approximate temperature $T_{H}$ on the Hubble horizon, in accordance with previous works \cite{Koma19,Koma20,Koma21}.
For this, we review a dynamical Kodama--Hayward temperature \cite{Dynamical-T-1998,Dynamical-T-2008}, which is interpreted as an extended version of the Gibbons--Hawking temperature  $T_{\rm{GH}}  = \frac{ \hbar H}{   2 \pi  k_{B}  }$ \cite{GibbonsHawking1977}.
The Kodama--Hayward temperature $T_{\rm{KH}}$ for a flat FLRW universe can be written as \cite{Tu2018,Tu2019}
\begin{equation}
 T_{\rm{KH}} = \frac{ \hbar H}{   2 \pi  k_{B}  }  \left ( 1 + \frac{ \dot{H} }{ 2 H^{2} }\right )  .
\label{eq:T_KH}
\end{equation}
Here $H>0$ and $\dot{H} \ge -2 H^{2}$ are assumed for a non-negative temperature in an expanding universe \cite{Koma19,Koma20,Koma21}.
In the present paper, the Kodama--Hayward temperature $T_{\rm{KH}}$ is used for the temperature $T_{H}$ on the horizon, to discuss the first law of thermodynamics in accordance with Refs.\ \cite{Cai2007,Cai2007B,Cai2008,Sanchez2023,Nojiri2024,Odintsov2023ab,Odintsov2024,Odintsov2024B,Nojiri2024B}.

\section{First law of thermodynamics and cosmological equations} 
\label{The first law}

In this section, we review the first law of thermodynamics and introduce cosmological equations derived from the first law, in accordance with Refs.\ \cite{Nojiri2024,Odintsov2024}.
Then, we reformulate the cosmological equations and determine the two extra driving terms, $f_{\Lambda}(t)$ and $h_{\textrm{B}}(t)$, based on the general formulation introduced in Sec.\ \ref{Cosmological equations}.
The first law of thermodynamics \cite{Cai2007,Cai2007B,Cai2008,Sanchez2023,Nojiri2024,Odintsov2023ab,Odintsov2024,Odintsov2024B,Nojiri2024B} was recently examined in a previous work \cite{Koma21} and, therefore, the first law is reviewed based on that work and the references therein.
Note that the Hubble horizon is equivalent to an apparent horizon because a flat FLRW universe is considered.

The first law of thermodynamics is written as \cite{Cai2007,Cai2007B,Cai2008,Sanchez2023,Nojiri2024,Odintsov2023ab,Odintsov2024,Odintsov2024B,Nojiri2024B}
\begin{align}
-dE_{\rm{bulk}}    + W dV  &  = T_{H} dS_{H}  ,
\label{eq:1stLaw}      
\end{align}
where $E_{\rm{bulk}}$ is the total internal energy of the matter fields inside the horizon, given by 
\begin{align}
E_{\rm{bulk}}  &= \rho c^{2} V .
\label{Ebulk}
\end{align}
$W$ represents the work density done by the matter fields \cite{Nojiri2024}, which is written as
\begin{align}
W  &= \frac{\rho c^{2} - p }{2}   ,
\label{eq:W1}      
\end{align}
and $V$ is the Hubble volume, written as
\begin{equation}
V = \frac{4 \pi}{3} r_{H}^{3} =  \frac{4 \pi}{3} \left ( \frac{c}{H} \right )^{3}   ,
\label{eq:V}
\end{equation}
where $r_{H} = c/H$ is given by Eq.\ (\ref{eq:rH}) \cite{Koma21}.
Equation\ (\ref{eq:1stLaw}) indicates that the entropy on the horizon is generated based on both the decreasing total internal energy of the bulk ($-dE_{\rm{bulk}}$) and the work done by the matter fields ($WdV$) \cite{Nojiri2024}.

In addition, Eq.\ (\ref{eq:1stLaw}) can be written as \cite{Koma21}
\begin{align}
- \frac{dE_{\rm{bulk}}}{dt}    + W \frac{dV}{dt}  &= T_{H} \frac{dS_{H}}{dt}  ,
\label{eq:1stLaw_dt0}      
\end{align}
or equivalently, 
\begin{align}
-\dot{E}_{\rm{bulk}}    + W \dot{V}  &=  T_{H} \dot{S}_{H}    .
\label{eq:1stLaw_dt1}      
\end{align}
Here the Kodama--Hayward temperature $T_{\rm{KH}}$ is used for the temperature $T_{H}$ on the horizon \cite{Cai2007,Cai2007B,Cai2008,Sanchez2023,Nojiri2024,Odintsov2023ab,Odintsov2024,Odintsov2024B,Nojiri2024B}.
In this study, an arbitrary entropy $S_{H}$ on the horizon is considered, as examined in Ref.\ \cite{Odintsov2024}.

To examine Eq.\ (\ref{eq:1stLaw_dt1}), we calculate the left side of this equation \cite{Koma21}.
Substituting Eqs.\ (\ref{Ebulk}) and  (\ref{eq:W1}) into $-\dot{E}_{\rm{bulk}}    + W \dot{V}$ yields \cite{Nojiri2024}
\begin{align}
-\dot{E}_{\rm{bulk}}    + W \dot{V}  &= - \frac{d (\rho c^{2} V)}{dt}                    + \left ( \frac{\rho c^{2} - p }{2} \right )  \dot{V}    \notag \\
                                                   &= -\dot{\rho} c^{2} V  - \left ( \frac{\rho c^{2} + p }{2} \right )  \dot{V}    .
\label{eq:Left1stLaw}      
\end{align}
Equation\ (\ref{eq:Left1stLaw}) corresponds to the left side of Eq.\ (\ref{eq:1stLaw_dt1}).
Therefore, Eq.\ (\ref{eq:1stLaw_dt1}) can be written as \cite{Odintsov2024}
\begin{align}
  -\dot{\rho} c^{2} V  - \left ( \frac{\rho c^{2} + p }{2} \right )  \dot{V}   &=  T_{H} \dot{S}_{H}     .
\label{eq:1stLaw_dt1_TdS}      
\end{align}
To derive cosmological equations, the standard continuity equation is applied \cite{Cai2007,Cai2007B,Cai2008,Sanchez2023,Nojiri2024,Odintsov2023ab,Odintsov2024,Odintsov2024B,Nojiri2024B}.
From Eq.\ (\ref{eq:drho_Zero}), the continuity equation is written as 
\begin{align}
       \dot{\rho} + 3  H \left ( \rho +  \frac{p}{c^2} \right )             &=0      ,
\label{eq:drho_General_Zero}
\end{align}
and $h_{\textrm{B}}(t)$ given by Eq.\ (\ref{eq:zero_hB_fL_org}) is written as 
\begin{equation}
       h_{\textrm{B}}(t)     =   \frac{  \dot{f}_{\Lambda}(t) }{ 2 H}     . 
\label{eq:zero_hB_fL}
\end{equation}

Based on the above preparations, we are able to derive cosmological equations from the first law of thermodynamics.
In fact, Odintsov \textit{et al.} derived cosmological equations from the first law using an arbitrary entropy $S_{H}$ on the horizon \cite{Odintsov2024}. 
The cosmological equations are considered to be a generalized cosmological model.
The derivation is summarized in Appendix \ref{Derivation of cosmological equations}, and the results are used here.
From Eq.\ (\ref{eq:dSHdSBH_rho_Odin_0}), the Friedmann equation from the first law is written as
\begin{align}
 \int \left (  \frac{\partial {S}_{H}}{\partial S_{\rm{BH}}} \right ) d (H^2)  &=   \frac{ 8 \pi G}{3}  \rho + C ,
\label{eq:dSHdSBH_rho_Odin}      
\end{align}
where $C$ is an integral constant and should be given by $\Lambda/3$.
The above equation corresponds to the Friedmann equation given by Eq.\ (\ref{eq:General_FRW01}).
In accordance with Ref.\ \cite{Odintsov2024}, we use a symbol with brackets, $(\partial S_{H}/\partial S_{\rm{BH}})$.
Note that the integral constant $C$ is retained here, to avoid confusion with $\Lambda$ measured by observations.
In addition, from Eq.\ (\ref{eq:dSHdSBH_rho-p_0}), a cosmological equation corresponding to Eq.\ (\ref{eq:dotH}) is written as 
\begin{align}
   \dot{H} \left (  \frac{\partial {S}_{H}}{\partial S_{\rm{BH}}} \right ) &=  - 4 \pi G  \left ( \rho + \frac{p}{c^{2}} \right ) .
\label{eq:dSHdSBH_rho-p}      
\end{align}
The acceleration equation is obtained from Eqs.\ (\ref{eq:dSHdSBH_rho_Odin}) and (\ref{eq:dSHdSBH_rho-p}), as examined later.
In fact, Eqs.\ (\ref{eq:dSHdSBH_rho_Odin}) and (\ref{eq:dSHdSBH_rho-p}) implicitly include two extra driving terms, namely $f_{\Lambda}(t)$ and $h_{\textrm{B}}(t)$, which are used for the general formulation introduced in Sec.\ \ref{Cosmological equations}.
However, it is difficult to find the two terms from the above two equations directly.
Therefore, in the next subsection, the cosmological equations are systematically formulated again, based on the general formulation.

\subsection{Reformulation of cosmological equations} 
\label{Reformulation of cosmological equations}

In this subsection, we reformulate the cosmological equations derived from the first law so that we can systematically determine $f_{\Lambda}(t)$ and $h_{\textrm{B}}(t)$ based on the general formulation.
To this end, the horizon entropy $S_{H}$ is reformulated in accordance with the work of Nojiri \textit{et al.} \cite{Nojiri2024}.
The horizon entropy $S_{H}$ can be written as
\begin{align}
S_{H} = S_{\rm{BH}} +  S_{\Delta}   ,
\label{eq:SH_SBH_dS}      
\end{align}
where $S_{\Delta}$ represents a deviation from the Bekenstein--Hawking entropy $S_{\rm{BH}}$.
Using this equation, $(\partial {S}_{H}/\partial S_{\rm{BH}})$ is written as
\begin{align}
\left (  \frac{\partial {S}_{H}}{\partial S_{\rm{BH}}} \right ) = 1 + \left ( \frac{\partial S_{\Delta} }{\partial S_{\rm{BH}}} \right )  .
\label{eq:dSHdSBH_1-dS}      
\end{align}
Substituting Eq.\ (\ref{eq:dSHdSBH_1-dS}) into the left side of Eq.\ (\ref{eq:dSHdSBH_rho_Odin}) yields
\begin{align}
 \int \left (  \frac{\partial {S}_{H}}{\partial S_{\rm{BH}}} \right ) d (H^2)  &=   \int \left [ 1 + \left (   \frac{\partial S_{\Delta} }{\partial S_{\rm{BH}}}  \right ) \right ] d (H^2)     \notag \\
                                                                                      &=  H^2 + \int \left ( \frac{\partial S_{\Delta} }{\partial S_{\rm{BH}}}  \right ) d (H^2)   .
\label{eq:Left_dSHdSBH_rho_Odin}      
\end{align}
Substituting the above equation into Eq.\ (\ref{eq:dSHdSBH_rho_Odin}) yields
\begin{align}
                       H^2 &=   \frac{ 8 \pi G}{3}  \rho + \underbrace{ C   - \int \left ( \frac{\partial S_{\Delta} }{\partial S_{\rm{BH}}}  \right ) d (H^2)  }_{f_{\Lambda}(t)} .
\label{eq:Fried_1st}      
\end{align}
Equation\ (\ref{eq:Fried_1st}) is the Friedmann equation and is equivalent to Eq.\ (\ref{eq:dSHdSBH_rho_Odin}).
The second and third terms on the right side of Eq.\ (\ref{eq:Fried_1st}) correspond to $f_{\Lambda}(t)$ in Eq.\ (\ref{eq:General_FRW01}).
When $S_{\Delta}=0$ (namely, $S_{H}=S_{\rm{BH}}$) is considered, Eq.\ (\ref{eq:Fried_1st}) reduces to $H^2 =   \frac{ 8 \pi G}{3}  \rho + C$.

Next, we formulate the acceleration equation.
Substituting Eq.\ (\ref{eq:SH_SBH_dS}) into Eq.\ (\ref{eq:dSHdSBH_rho-p}) yields
\begin{align}
   \dot{H} \left (  \frac{\partial {S}_{H}}{\partial S_{\rm{BH}}} \right ) &= \dot{H} \left [ 1 + \left (   \frac{\partial S_{\Delta} }{\partial S_{\rm{BH}}} \right ) \right ]   = - 4 \pi G  \left ( \rho + \frac{p}{c^{2}} \right )  .
\end{align}
This equation can be written as
\begin{align}
      \dot{H}  &=  - 4 \pi G  \left ( \rho + \frac{p}{c^{2}} \right )     \underbrace{  -    \dot{H} \left ( \frac{\partial S_{\Delta} }{\partial S_{\rm{BH}}}  \right )  }_{h_{\textrm{B}}(t)}     .
\label{eq:dotH_1}      
\end{align}
The second term on the right side of Eq.\ (\ref{eq:dotH_1}) corresponds to $h_{\textrm{B}}(t)$ in Eq.\ (\ref{eq:dotH}).
Substituting Eqs.\  (\ref{eq:dotH_1}) and (\ref{eq:Fried_1st}) into $\ddot{a}/a =\dot{H} + H^{2}$ yields 
\begin{widetext}
\begin{equation}
  \frac{ \ddot{a}}{a}   = \dot{H} + H^{2}          
                                = -  \frac{ 4\pi G }{ 3 }  \left ( \rho +  \frac{3 p}{c^2} \right )   
                                   + \underbrace{ C   - \int \left ( \frac{\partial S_{\Delta} }{\partial S_{\rm{BH}}}  \right ) d (H^2)  }_{f_{\Lambda}(t)}   \underbrace{  -    \dot{H} \left ( \frac{\partial S_{\Delta} }{\partial S_{\rm{BH}}}  \right )  }_{h_{\textrm{B}}(t)}     .
\label{eq:FRW02_1st}
\end{equation}
\end{widetext}
Equation\ (\ref{eq:FRW02_1st}) is the acceleration equation derived from the first law of thermodynamics.
When $S_{\Delta}=0$ (i.e., $S_{H}=S_{\rm{BH}}$), Eq.\ (\ref{eq:FRW02_1st}) reduces to $\frac{\ddot{a}}{a} =   -  \frac{ 4\pi G }{ 3 }  \left ( \rho +  \frac{3 p}{c^2} \right )  + C$.

From Eqs.\ (\ref{eq:Fried_1st}), (\ref{eq:FRW02_1st}), and (\ref{eq:drho_General_Zero}), the Friedmann, acceleration, and continuity equations for the present model are written as
\begin{equation}
 H(t)^2      =  \frac{ 8\pi G }{ 3 } \rho (t)    + f_{\Lambda}(t)            ,                                                 
\label{eq:General_FRW01_1st} 
\end{equation} 
\begin{align}
  \frac{ \ddot{a}}{a}    &= -  \frac{ 4\pi G }{ 3 }  \left ( \rho +  \frac{3 p}{c^2} \right )                   +   f_{\Lambda}(t)    +  h_{\textrm{B}}(t)  , 
\label{eq:General_FRW02_1st}
\end{align}
\begin{align}
 \dot{\rho} + 3  H \left ( \rho +  \frac{p}{c^2} \right )  &=0 ,
\end{align}
where the two extra driving terms, $f_{\Lambda}(t)$ and $h_{\textrm{B}}(t)$, are given by
\begin{align}
                        f_{\Lambda}(t)  &=  C   - \int \left ( \frac{\partial S_{\Delta} }{\partial S_{\rm{BH}}}  \right ) d (H^2)   ,
\label{eq:fL(t)_1stLaw}      
\end{align}
\begin{align}
                         h_{\textrm{B}}(t)  &=   -    \dot{H} \left ( \frac{\partial S_{\Delta} }{\partial S_{\rm{BH}}}  \right )      .
\label{eq:hB(t)_1stLaw}      
\end{align}
The two terms include $(\partial S_{\Delta} / \partial S_{\rm{BH}})$.
Using $d(H^2)= 2HdH$, we can confirm that Eqs.\ (\ref{eq:fL(t)_1stLaw}) and (\ref{eq:hB(t)_1stLaw}) satisfy Eq.\ (\ref{eq:zero_hB_fL}), namely $h_{\textrm{B}}(t)   =   \dot{f}_{\Lambda}(t) / (2H)$.
When $S_{\Delta}=0$, the two terms reduce to $f_{\Lambda}(t) = C$ and $h_{\textrm{B}}(t) =0$, respectively.
Accordingly, the cosmological equations for the present model for $S_{\Delta}=0$ are equivalent to those for the $\Lambda$CDM models, although the theoretical backgrounds are different.
Various forms of entropy \cite{Das2008,Radicella2010,Tsallis2012,Czinner1Czinner2,Barrow2020,Nojiri2022} can be applied to the present model because an arbitrary entropy $S_{H}$ is considered.
(In Appendix\ \ref{The present model for a power-law corrected entropy}, a power-law-corrected entropy \cite{Das2008,Radicella2010} is applied to the present model.)

In this section, we derive cosmological equations from the first law of thermodynamics using an arbitrary entropy $S_{H}$ on the horizon, in accordance with Refs.\ \cite{Nojiri2024,Odintsov2024}.
In addition, we formulate the cosmological equations again and determine the two extra driving terms, $f_{\Lambda}(t)$ and $h_{\textrm{B}}(t)$, based on the general formulation.
That is, we formulate a generalized cosmological model derived from the first law.
The present model expresses the two driving terms explicitly.
We expect that $(\partial S_{\Delta} / \partial S_{\rm{BH}})$ included in the two terms plays important roles in the discussion of thermodynamic constraints.
In the next section, the thermodynamic constraints are universally examined.

\section{Second law of thermodynamics and thermodynamic constraints on the present model} 
\label{Thermodynamic constraints}

In the previous section, we formulated a generalized cosmological model from the first law of thermodynamics and determined two extra driving terms, $f_{\Lambda}(t)$ and $h_{\textrm{B}}(t)$.
From Eqs.\ (\ref{eq:fL(t)_1stLaw}) and (\ref{eq:hB(t)_1stLaw}), the two terms for the present model are written as
\begin{align}
                        f_{\Lambda}(t)  &=  C   - \int \left ( \frac{\partial S_{\Delta} }{\partial S_{\rm{BH}}}  \right ) d (H^2)   ,
\label{eq:fL(t)_1stLaw_2}      
\end{align}
\begin{align}
                         h_{\textrm{B}}(t)  &=   -    \dot{H} \left ( \frac{\partial S_{\Delta} }{\partial S_{\rm{BH}}}  \right )      ,
\label{eq:hB(t)_1stLaw_2}      
\end{align}
where the deviation $S_{\Delta}$ is $S_{H} - S_{\rm{BH}}$, which is given by Eq.\ (\ref{eq:SH_SBH_dS}).
Note that $S_{H}$ is an arbitrary entropy on the horizon and $S_{\rm{BH}}$ is the Bekenstein--Hawking entropy.

In this section, based on the second law of thermodynamics, we universally examine thermodynamic constraints on the two terms in the present model, extending the method used in previous works \cite{Koma10,Koma11,Koma12}.
Similar constraints are discussed in those works, using holographic equipartition models that are different from the present model.
(The second law itself has been examined: for example, to discuss the second law, the total entropy, namely the sum of the horizon entropy and the entropy of the matter fields, was considered \cite{Odintsov2024}.)
In our Universe, the horizon entropy is extremely larger than the sum of the other entropies \cite{Egan1}.
In addition, the horizon entropy is included in the two terms.
Therefore, we focus on the horizon entropy.
Consequently, the second law of thermodynamics is written as
\begin{align}
                 \dot{S}_{H} \ge 0   . 
\label{eq:2ndLaw_SH}      
\end{align}
As examined in Sec.\ \ref{Entropy and temperature}, we consider $\dot{S}_{\rm{BH}} >  0$ given by Eq.\ (\ref{eq:dSBH_2}), because $H >0$ and $\dot{H} < 0$ in our Universe \cite{Hubble2017}.
Also, $\dot{H} \ge -2 H^{2}$ is assumed because the horizon temperature given by Eq.\ (\ref{eq:T_KH}) is considered to be non-negative.
In addition, $S_{\Delta} \neq 0$ is considered.
Note that various nonextensive entropies satisfy $S_{\Delta} \neq 0$, except when $S_{H} = S_{\rm{BH}}$ \cite{Das2008,Radicella2010,Tsallis2012,Czinner1Czinner2,Barrow2020,Nojiri2022}.

We now examine thermodynamic constraints on the present model.
Substituting Eq.\ (\ref{eq:dSHdSBH_1-dS}) into Eq.\ (\ref{eq:hB(t)_1stLaw_2}) yields
\begin{align}
                         h_{\textrm{B}}(t)  &=   - \dot{H} \left ( \frac{\partial S_{\Delta} }{\partial S_{\rm{BH}}}  \right )   =      \dot{H} \left [ 1 -  \left ( \frac{\partial S_{H} }{\partial S_{\rm{BH}}} \right )  \right ]  \notag \\
                                                   &=     \dot{H} \left ( 1 -   \frac{\dot{S}_{H} }{\dot{S}_{\rm{BH}}}  \right ) ,
\label{eq:hB(t)_1stLaw_reform}      
\end{align}
where $(\partial S_{H}/ \partial S_{\rm{BH}}) = \dot{S}_{H} / \dot{S}_{\rm{BH}}$ is used from Eq.\ (\ref{eq:dSH_reform_dSBH}).
Solving Eq.\ (\ref{eq:hB(t)_1stLaw_reform}) with regard to $\dot{S}_{H}$ gives
\begin{align}
                         \dot{S}_{H}   &=   \dot{S}_{\rm{BH}} \left ( 1 -   \frac{h_{\textrm{B}}(t)}{\dot{H}}  \right ) .
\label{eq:dSHdt_hB(t)}      
\end{align}
From Eq.\ (\ref{eq:dSHdt_hB(t)}), to satisfy $\dot{S}_{H} \ge  0 $, we require
\begin{equation}
  \frac{h_{\textrm{B}}(t)}{\dot{H}}  \le 1      ,
\label{dSHdt_ineq_hB_1}      
\end{equation}
where $\dot{S}_{\rm{BH}} >  0$ is used. 
This inequality corresponds to the thermodynamic constraint on $h_{\textrm{B}}(t)$.
Applying $\dot{H}< 0$ to Eq.\ (\ref{dSHdt_ineq_hB_1}) gives 
\begin{equation}
  h_{\textrm{B}}(t)  \ge \dot{H}   \quad (\textrm{for} \quad  \dot{H} < 0)   .
\label{dSHdt_ineq_hB_Negative-dH}      
\end{equation}
The above inequality implies a lower limit of $h_{\textrm{B}}(t)$.
The lower limit is negative because $\dot{H} <0$.
Substituting Eq.\ (\ref{eq:dotH}) into Eq.\ (\ref{dSHdt_ineq_hB_Negative-dH}) yields
\begin{equation}
\dot{H} + 4\pi G  \left ( \rho +  \frac{p}{c^2} \right )  \ge \dot{H}   \quad (\textrm{for} \quad  \dot{H} < 0)   .
\label{dSHdt_ineq_hB_Negative-dH_A1}      
\end{equation}
In addition, substituting $w=p/(\rho c^{2})$ into Eq.\ (\ref{dSHdt_ineq_hB_Negative-dH_A1}) gives
\begin{equation}
     1+w \ge 0 \quad (\textrm{for} \quad  \dot{H} < 0)   ,
\label{dSHdt_ineq_hB_Negative-dH_A2}      
\end{equation}
where $\rho$ is positive. 
In this way, the constraint on $w$ can be obtained from the constraint on $h_{\textrm{B}}(t)$.
We note that $w > -1$ considered in Sec.\ \ref{Cosmological equations} satisfies Eq.\ (\ref{dSHdt_ineq_hB_Negative-dH_A2}).
An upper limit of $h_{\textrm{B}}(t)$ and the order of $h_{\textrm{B}}(t)$ are discussed later.

In the present study, $h_{\textrm{B}}(t)$ is given by Eq.\ (\ref{eq:zero_hB_fL}), namely $\dot{f}_{\Lambda}(t) / (2H)$, because the standard continuity equation is considered. 
Substituting Eq.\ (\ref{eq:zero_hB_fL}) into Eq.\ (\ref{dSHdt_ineq_hB_1}) yields a constraint on $\dot{f}_{\Lambda}(t)$:
\begin{equation}
      \frac{\dot{f}_{\Lambda}(t) / (2H)}{\dot{H}}   = \frac{\dot{f}_{\Lambda}(t)}{2 H \dot{H}}   \le 1      .
\label{dSHdt_ineq__fL_mod}      
\end{equation}
Also, from Eq.\ (\ref{eq:fL(t)_1stLaw_2}), we can obtain a similar constraint on $df_{\Lambda}(t)$:
\begin{equation}
 \frac{d f_{\Lambda}(t)}{d (H^2)}   \le 1      .
\label{dSHdt_ineq_dfL}      
\end{equation}
The two inequalities correspond to the constraints on $\dot{f}_{\Lambda}(t)$ and $df_{\Lambda}(t)$.
However, the two inequalities do not constrain the extent of $f_{\Lambda}(t)$, although the two should help to examine cosmological models from a thermodynamics viewpoint.

In fact, thermodynamic constraints on $f_{\Lambda}(t)$ can be discussed using Eq.\ (\ref{eq:Back2}), which is written as
\begin{equation}
    \dot{H}       = -  \frac{3}{2} (1+w)  H^{2} \left ( 1- \frac{f_{\Lambda}(t)}{H^{2}} \right )    + h_{\textrm{B}}(t)   ,
\label{eq:Back2_2}
\end{equation}
where $w > -1$ is considered, as examined in Sec.\ \ref{Cosmological equations}.
Solving Eq.\ (\ref{eq:Back2_2}) with regard to $h_{\textrm{B}}(t)$ and substituting the resultant equation into Eq.\ (\ref{eq:dSHdt_hB(t)}) yields 
\begin{align}
                         \dot{S}_{H}   &=   \dot{S}_{\rm{BH}} \left ( 1 -   \frac{h_{\textrm{B}}(t)}{\dot{H}}  \right )  \notag \\
                                           &= \dot{S}_{\rm{BH}} \left ( 1 -   \frac{\dot{H} + \frac{3}{2} (1+w)  H^{2} \left ( 1- \frac{f_{\Lambda}(t)}{H^{2}} \right ) }{\dot{H}}  \right ) \notag \\
                                           &= \dot{S}_{\rm{BH}} \left [ \frac{3}{2} (1+w) \left( - \frac{ H^{2}}{\dot{H}} \right ) \left ( 1- \frac{f_{\Lambda}(t)}{H^{2}} \right )  \right ]   .
\label{eq:dSHdt_hB(t)_fL(t)}      
\end{align}
Using $\dot{S}_{\rm{BH}} >  0$, $\dot{H} < 0$, and $w > -1$ to satisfy $\dot{S}_{H} \ge  0 $, we require
\begin{equation}
   1- \frac{f_{\Lambda}(t)}{H^{2}}   \ge 0      ,
\label{dSHdt_ineq_fL_0}      
\end{equation}
or equivalently,
\begin{equation}
   f_{\Lambda}(t)  \le H^{2}      .
\label{dSHdt_ineq_fL_1}      
\end{equation}
Equations\ (\ref{dSHdt_ineq_fL_0}) and (\ref{dSHdt_ineq_fL_1}) imply an upper limit of $f_{\Lambda}(t)$.
When  $0< H$ and $H_{0} \leq H$ (obtained from $\dot{H} < 0$), the strictest constraint from the past to the present is given by
\begin{equation}
f_{\Lambda}(t)     \leq   H_{0}^{2}     \leq H^{2}     ,
\label{dSHdt_ineq_fL_H0_1}
\end{equation}
and the order of $f_{\Lambda}(t)$ can be written as
\begin{equation}
  O  ( f_{\Lambda}(t) )  \lessapprox     O  (  H_{0}^{2}  )       .
\label{dSHdt_ineq_fL_order}
\end{equation}
In addition, we can examine an upper limit of $h_{\textrm{B}}(t)$.
Applying $\dot{H} < 0$ to Eq.\ (\ref{eq:Back2_2}) and using Eq.\ (\ref{dSHdt_ineq_fL_1}) yields 
\begin{align}
                 h_{\textrm{B}}(t)   &= \dot{H} + \frac{3}{2} (1+w)  H^{2} \left ( 1- \frac{f_{\Lambda}(t)}{H^{2}} \right ) \notag \\
                                            &\leq \frac{3}{2} (1+w)  H^{2} ,
\label{eq:hB(t)_upper}      
\end{align}
where $f_{\Lambda}(t) \geq 0$ is also used.
(Such a non-negative $f_{\Lambda}(t)$ can be obtained from, e.g., a power-law-corrected entropy, as examined in Appendix \ref{The present model for a power-law corrected entropy}.)
Equation\ (\ref{eq:hB(t)_upper}) implies the upper limit of $h_{\textrm{B}}(t)$.
Coupling Eq.\ (\ref{dSHdt_ineq_hB_Negative-dH}) with Eq.\ (\ref{eq:hB(t)_upper}) and applying $\dot{H} \ge -2 H^{2}$ yields 
\begin{equation}
  -2 H^{2} \leq \dot{H} \leq h_{\textrm{B}}(t)  \leq \frac{3}{2} (1+w)  H^{2}  \quad (\textrm{for} \quad  \dot{H} < 0)   .
\label{hB_lower-upper}      
\end{equation}
This inequality corresponds to an extended thermodynamic constraint on $h_{\textrm{B}}(t)$.
Applying $0< H_{0} \leq H$ to Eq.\ (\ref{hB_lower-upper}) gives the order of $h_{\textrm{B}}(t)$, written as
\begin{equation}
O  ( - H_{0}^{2}  )    \lessapprox       O  ( h_{\textrm{B}}(t)  )  \lessapprox     O  (  H_{0}^{2}  )     .
\label{dSHdt_ineq_hB(t)_order}
\end{equation}
These results indicate that the second law of thermodynamics should constrain $f_{\Lambda}(t)$ and $h_{\textrm{B}}(t)$.
From Eqs.\ (\ref{dSHdt_ineq_fL_order}) and (\ref{dSHdt_ineq_hB(t)_order}), the orders of the upper limits of $f_{\Lambda}(t)$ and $h_{\textrm{B}}(t)$ are likely consistent with the order of the cosmological constant measured by observations, namely $\Lambda_{\textrm{obs}}$.
We note that $O(\Lambda_{\textrm{obs}}/3) \approx O(H_{0}^{2})$ is given by Eq.\ (\ref{L_order_0}), which is based on the Planck 2018 results \cite{Planck2018}.
Also, from Eq.\ (\ref{dSHdt_ineq_hB(t)_order}), the absolute value of the order of the lower limit of $h_{\textrm{B}}(t)$ is the same as the order of the upper limit.

\begin{figure} [t]  
\begin{minipage}{0.49\textwidth}
\begin{center}
\scalebox{0.32}{\includegraphics{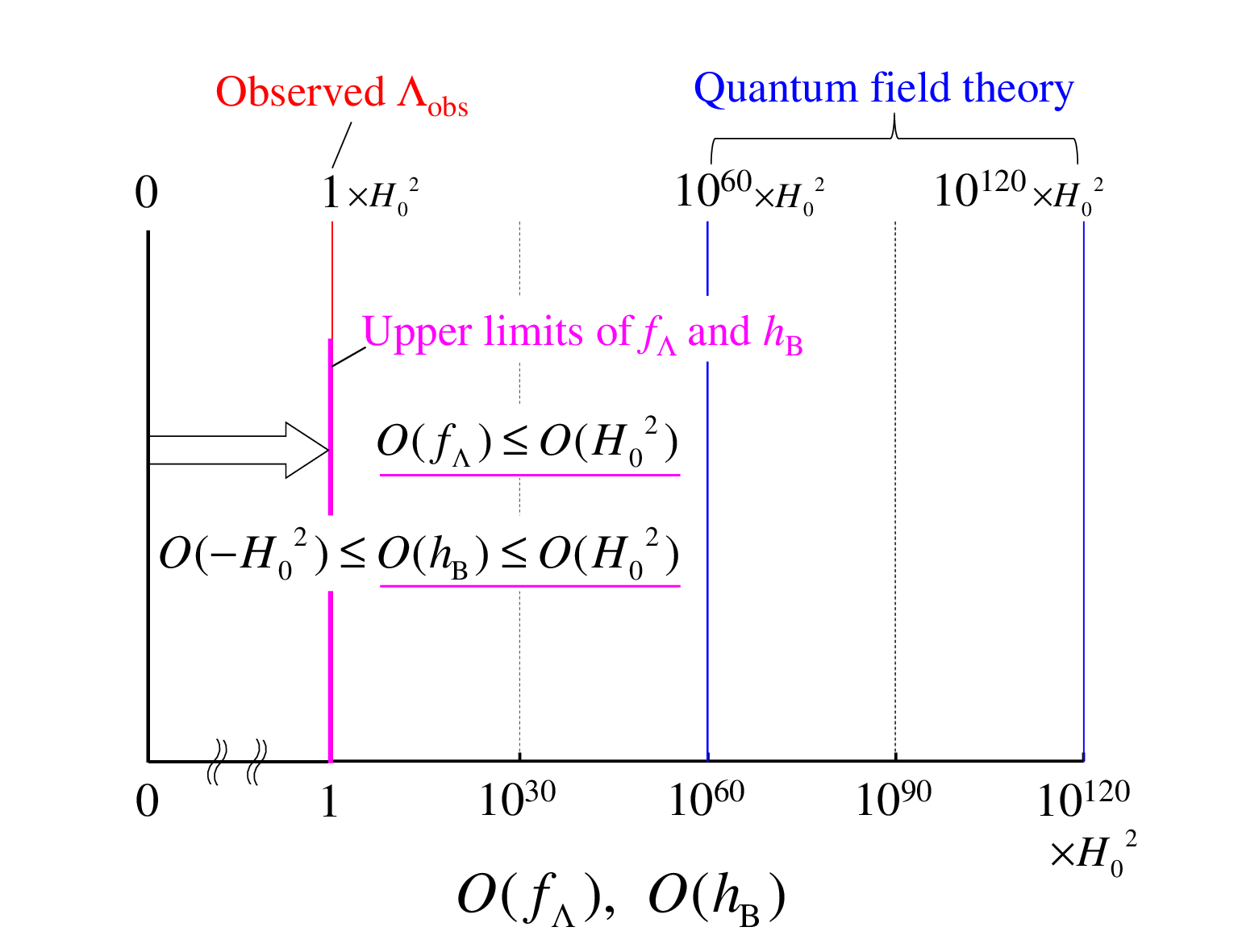}}
\end{center}
\end{minipage}
\caption{Thermodynamic constraints on the two driving terms for the present model.
The constraints on $f_{\Lambda}(t)$ and $h_{\textrm{B}}(t)$ are given by Eqs.\ (\ref{dSHdt_ineq_fL_order}) and (\ref{dSHdt_ineq_hB(t)_order}), respectively.
An arrow represents an allowed region for the upper limits of the two terms.
In Refs.\ \cite{Koma10,Koma11,Koma12}, similar constraints on an extra driving term are discussed using holographic equipartition models.
}
\label{Fig1r}
\end{figure}

Figure\ \ref{Fig1r} shows the orders of the upper limits of the two terms, the order of $\Lambda_{\textrm{obs}}$ \cite{Planck2018}, and the order of the theoretical value from quantum field theory \cite{Weinberg1989,Pad2003,Barrow2011,Bao2017}.
A region for the lower limit of $h_{\textrm{B}}(t)$ is not shown in this figure because the lower limit is negative and the order is the same as the order of the upper limit.
We can confirm that the orders of the upper limits of the two terms are consistent with the order of $\Lambda_{\textrm{obs}}$.
The discrepancy of $60 \sim 120$ orders of magnitude appears to be avoided, as if a consistent scenario could be obtained from thermodynamics.

Finally, we use the present model to discuss an important model similar to the $\Lambda$CDM models ($f_{\Lambda}(t) \rightarrow C$ and $h_{\textrm{B}}(t) \rightarrow  0$).
To this end, a deviation $S_{\Delta}$ from the Bekenstein--Hawking entropy is considered to be close to zero but not zero.
Accordingly, Eqs.\ (\ref{eq:fL(t)_1stLaw_2}) and (\ref{eq:hB(t)_1stLaw_2}) should reduce to $f_{\Lambda}(t) \approx C + \epsilon_{1}$ and $h_{\textrm{B}}(t) \approx \epsilon_{2}$, respectively.
Here, two parameters, $ \epsilon_{1}$ and $\epsilon_{2}$, should be close to zero because $S_{\Delta}$ is close to zero.
Therefore, the constraint on $h_{\textrm{B}}(t)$ given by Eq.\ (\ref{dSHdt_ineq_hB_1}) should be satisfied, and the order of $h_{\textrm{B}}(t)$ can be given by
\begin{equation}
 O [h_{\textrm{B}}(t) ] \approx  O  ( \epsilon_{2} ) \approx 0    .
\label{dSHdt_ineq_hB-C_order}
\end{equation}
In addition, the constraint on $f_{\Lambda}(t)$ given by Eq.\ (\ref{dSHdt_ineq_fL_order}) can be written as
\begin{equation}
 O  ( C )  \approx  O  ( C + \epsilon_{1} )  \approx  O  [ f_{\Lambda}(t) ]   \lessapprox     O  (  H_{0}^{2} )   .
\label{dSHdt_ineq_fL-C_order}
\end{equation}
Equation\ (\ref{dSHdt_ineq_fL-C_order}) implies that the order of $C$ is consistent with the order of $\Lambda_{\textrm{obs}}$.
In this sense, we can use the present model to discuss the cosmological constant problem from a thermodynamic viewpoint when $S_{\Delta}$ is close to zero but not zero.
The small deviation $S_{\Delta}$ from the Bekenstein--Hawking entropy may play an important role in the accelerated expansion of our late Universe. 

An arbitrary entropy $S_{H}$ on the horizon is considered here and, therefore, we can use various forms of entropy on the horizon; see, for example, Appendix \ref{The present model for a power-law corrected entropy} and Refs.\ \cite{Odintsov2024,Nojiri2024B}.
Before fine-tuning, we should be able to discuss thermodynamic constraints on driving terms calculated from those entropies.
We expect that effective entropies, which deviate slightly from the Bekenstein--Hawking entropy, are favored in our Universe.
Further studies are needed, and those tasks are left for future research.

It should be noted that, in this section, we consider $S_{\Delta} \neq 0$ and discuss the thermodynamic constraints. 
When $S_{\Delta}=0$ (i.e., $S_{H}=S_{\rm{BH}}$), the two terms reduce to $f_{\Lambda}(t) = C$ and $h_{\textrm{B}}(t) =0$, respectively, as for $\Lambda$CDM models.
In this case, Eq.\ (\ref{eq:dSHdt_hB(t)}) reduces to $\dot{S}_{H}=\dot{S}_{\rm{BH}}$.
In fact, Eq.\ (\ref{eq:dSHdt_hB(t)}) is used in the calculation of the constraint on $f_{\Lambda}(t)$, through Eq.\ (\ref{eq:dSHdt_hB(t)_fL(t)}).
Accordingly, when $S_{\Delta}=0$, it is difficult to discuss the thermodynamic constraints.
To avoid this difficulty, we consider $S_{\Delta} \neq 0$.
Various nonextensive entropies satisfy $S_{\Delta} \neq 0$, except when $S_{H} = S_{\rm{BH}}$.
Note that we also consider $\dot{S}_{\rm{BH}} >  0$, $H > 0$, $-2 H^{2} \le \dot{H} < 0$, and $w > -1$.

In this paper, we examine the thermodynamic constraints on the two driving terms in a generalized cosmological model derived from the first law of thermodynamics, using an arbitrary entropy on the horizon.
The thermodynamic constraints imply that the orders of the two terms are consistent with the order of the cosmological constant measured by observations.
Of course, we cannot exclude all the other contributions, such as quantum field theory, because these contributions have not been examined in this study.
In addition, the assumptions used here have not yet been established but are considered to be viable.
The present study should contribute to a better understanding of thermodynamic scenarios, which may provide new insights into the discussion of the cosmological constant problem.

\section{Conclusions}
\label{Conclusions}

The first and second laws of thermodynamics should lead to a consistent scenario for discussing the cosmological constant problem.
To establish such a thermodynamic scenario, we have derived cosmological equations in a flat FLRW universe from the first law, using an arbitrary entropy $S_{H}$ on the horizon.
The derived cosmological equations implicitly include extra driving terms. 
Therefore, we have systematically reformulated the cosmological equations using a general formulation that includes two extra driving terms, $f_{\Lambda}(t)$ and $h_{\textrm{B}}(t)$, explicitly.
The present model is essentially equivalent to that derived in Ref.\ \cite{Odintsov2024}, but is suitable for the discussion of thermodynamic constraints. 

Based on the second law of thermodynamics, we have universally examined the thermodynamic constraints on the two terms in the present model.
It is found that a variation in the deviation $S_{\Delta}$ of $S_{H}$ from the Bekenstein--Hawking entropy plays an important role in the thermodynamic constraints.
In our late Universe, the second law should constrain both the upper limit of $f_{\Lambda}(t)$ and the upper and lower limits of $h_{\textrm{B}}(t)$.
The upper limits imply that the orders of the two terms are consistent with the order of the cosmological constant $\Lambda_{\textrm{obs}}$ measured by observations.
The lower limit of $h_{\textrm{B}}(t)$ leads to a constraint on $w$, and the absolute value of the order of the lower limit is likely consistent with the order of $\Lambda_{\textrm{obs}}$ as well.
In particular, when the deviation $S_{\Delta}$ is close to zero but not zero, $h_{\textrm{B}}(t)$ and $f_{\Lambda}(t)$ should reduce to zero and a constant (consistent with the order of $\Lambda_{\textrm{obs}}$), respectively, as if a consistent and viable scenario could be obtained from thermodynamics.
The thermodynamic scenario may imply that the accelerated expansion of our Universe is related to the small deviation from the Bekenstein--Hawking entropy because the deviation included in the two terms can also contribute to the accelerated expansion.

\appendix

\section{Derivation of cosmological equations} 
\label{Derivation of cosmological equations}

In this appendix, we derive cosmological equations from the first law of thermodynamics based on the work of Odintsov \textit{et al.} \cite{Odintsov2024}.
The cosmological equations are considered to constitute a generalized cosmological model derived from the first law.
Substituting $\dot{\rho} = - 3  H \left ( \rho +  \frac{p}{c^2} \right )$ given by Eq.\ (\ref{eq:drho_General_Zero}) into Eq.\ (\ref{eq:1stLaw_dt1_TdS}) yields 
\begin{align}
  T_{H} \dot{S}_{H}   &= -\dot{\rho} c^{2} V  - \left ( \frac{\rho c^{2} + p }{2} \right )  \dot{V}  \notag \\
                            &=  3  H \left ( \rho +  \frac{p}{c^2} \right ) c^{2}  V - \left ( \frac{\rho c^{2} + p }{2} \right ) \dot{V}  \notag \\
                            &= ( \rho c^{2} + p ) \left (  3HV - \frac{\dot{V} }{2} \right )  .
\label{eq:1stLaw_dt1_reform}      
\end{align}
Solving Eq.\ (\ref{eq:1stLaw_dt1_reform}) with regard to $\dot{S}_{H}$, substituting $V= (4 \pi/3)(c/H)^{3}$ given by Eq.\ (\ref{eq:V}) and $\dot{V}= -4 \pi c^{3} H^{-4} \dot{H}$ into the resultant equation, and performing several operations yields \cite{Odintsov2024}
\begin{align}
   \dot{S}_{H}  &=  \frac{ ( \rho c^{2} + p ) \left (  3HV - \frac{\dot{V} }{2} \right )  }{   T_{H}   }     \notag \\
                           &= \frac{ ( \rho c^{2} + p )  \left (  3H \frac{4 \pi}{3} \left ( \frac{c}{H} \right )^{3}  - \frac{(-4 \pi c^{3} H^{-4} \dot{H}) }{2} \right )       }{    \frac{ \hbar H}{   2 \pi  k_{B}  }  \left ( 1 + \frac{ \dot{H} }{ 2 H^{2} }\right )     }  \notag \\
                           &=  \left ( \rho + \frac{p}{c^{2}} \right )  \frac{ 8 \pi^{2} }{H^{3}} \frac{k_{B} c^{5}}{\hbar}   ,
\label{eq:dSH_reform}      
\end{align}
where $T_{\rm{KH}}$ given by Eq.\ (\ref{eq:T_KH}) is used for $T_{H}$.
Conversely, using the form of the Bekenstein--Hawking entropy $S_{\rm{BH}}$ allows us to write $\dot{S}_{H}$ as 
\begin{align}
   \dot{S}_{H}  &=  \left (  \frac{\partial {S}_{H}}{\partial S_{\rm{BH}}} \right ) \frac{\partial S_{\rm{BH}}}{\partial t}   =  \left (  \frac{\partial {S}_{H}}{\partial S_{\rm{BH}}} \right ) \frac{-2 K  \dot{H} }{H^{3}}    \notag \\
                    &=  \left (  \frac{\partial {S}_{H}}{\partial S_{\rm{BH}}} \right ) \frac{-2 \frac{  \pi  k_{B}  c^5 }{ \hbar G }  \dot{H} }{H^{3}}    ,
\label{eq:dSH_reform_dSBH}      
\end{align}
where $\dot{S}_{\rm{BH}} = \frac{-2K \dot{H} }{H^{3}}$ given by Eq.\ (\ref{eq:dSBH}) and $K = \frac{  \pi  k_{B}  c^5 }{ \hbar G }$ given by Eq.\ (\ref{eq:K-def}) are applied.
Based on Ref.\ \cite{Odintsov2024}, we use a symbol with brackets, namely $(\partial S_{H}/\partial S_{\rm{BH}})$.
To this end, in the above calculation, $\dot{S}_{H} = \partial S_{H}/ \partial t$ and $\dot{S}_{\rm{BH}} = \partial S_{\rm{BH}}/ \partial t$ are applied and, therefore, $(\partial S_{H}/\partial S_{\rm{BH}})$ corresponds to $\dot{S}_{H} / \dot{S}_{\rm{BH}}$.

Substituting Eq.\ (\ref{eq:dSH_reform}) into Eq.\ (\ref{eq:dSH_reform_dSBH}) yields 
\begin{align}
   \left ( \rho + \frac{p}{c^{2}} \right )  \frac{ 8 \pi^{2} }{H^{3}} \frac{k_{B} c^{5}}{\hbar} &= \left (  \frac{\partial {S}_{H}}{\partial S_{\rm{BH}}} \right ) \frac{-2 \frac{  \pi  k_{B}  c^5 }{ \hbar G }  \dot{H} }{H^{3}}  .
\label{eq:rho-p_dSHdSBH}      
\end{align}
Calculating this equation gives \cite{Odintsov2024}
\begin{align}
   \dot{H} \left (  \frac{\partial {S}_{H}}{\partial S_{\rm{BH}}} \right ) &=  - 4 \pi G  \left ( \rho + \frac{p}{c^{2}} \right ) .
\label{eq:dSHdSBH_rho-p_0}      
\end{align}
The above equation corresponds to Eq.\ (\ref{eq:dotH}).
Substituting $\dot{\rho} + 3  H \left ( \rho +  \frac{p}{c^2} \right )  =  0$ given by Eq.\ (\ref{eq:drho_General_Zero}) into Eq.\ (\ref{eq:dSHdSBH_rho-p_0}) yields 
\begin{align}
   \dot{H} \left (  \frac{\partial {S}_{H}}{\partial S_{\rm{BH}}} \right ) &=  - 4 \pi G  \left ( \rho + \frac{p}{c^{2}} \right ) =  4 \pi G    \left ( \frac{\dot{\rho}}{3H} \right ) .
\label{eq:dH-dSHdSBH_rho}      
\end{align}
We can rearrange Eq.\ (\ref{eq:dH-dSHdSBH_rho}) as
\begin{align}
 \left (  \frac{\partial {S}_{H}}{\partial S_{\rm{BH}}} \right )   H \dot{H} &=  \frac{ 4 \pi G}{3}  \dot{\rho}   ,
\label{eq:dH-dSHdSBH_drho-dt}      
\end{align}
or equivalently,
\begin{align}
\left (  \frac{\partial {S}_{H}}{\partial S_{\rm{BH}}} \right )   H dH  &=  \frac{ 4 \pi G}{3}  d \rho   .
\label{eq:dH-dSHdSBH_drho}      
\end{align}
Integrating Eq.\ (\ref{eq:dH-dSHdSBH_drho}) and applying $H dH = \frac{1}{2} d (H^2)$ yields 
\begin{align}
 \int \left (  \frac{\partial {S}_{H}}{\partial S_{\rm{BH}}} \right ) d (H^2)  &=   \frac{ 8 \pi G}{3}  \rho + C ,
\label{eq:dSHdSBH_rho_Odin_0}      
\end{align}
where $C$ is an integral constant and should be given by $C= \Lambda/3$.
Equation\ (\ref{eq:dSHdSBH_rho_Odin_0}) is the Friedmann equation derived from the first law of thermodynamics, which is examined in Ref.\ \cite{Odintsov2024}.
In Sec.\ \ref{The first law}, we reformulate those cosmological equations, based on the general formulation introduced in Sec.\ \ref{Cosmological equations}.

\section{Present model for a power-law-corrected entropy $S_{pl}$}
\label{The present model for a power-law corrected entropy}

In this appendix, as a specific entropy, we apply a power-law-corrected entropy to the present model.
The power-law-corrected entropy suggested by Das \textit{et al.} \cite{Das2008} is based on the entanglement of quantum fields between the inside and outside of the horizon and has been applied to holographic equipartition models \cite{Koma11,Koma12,Koma14}.
As far as we know, the power-law-corrected entropy has not yet been applied to the present model.
Note that several forms of entropy have been examined in, for example, the recent review by Nojiri \textit{et al.} \cite{Nojiri2024B} and the references therein.

The power-law-corrected entropy $S_{pl}$ \cite{Radicella2010} can be written as  
\begin{equation}
 S_{pl}  = S_{\rm{BH}}  \left [ 1-  \psi_{\alpha} \left ( \frac{H_{0}}{H} \right )^{2- \alpha}  \right ] , 
\label{eq:Spl}     
\end{equation}
where $\psi_{\alpha}$ is a dimensionless parameter given by 
\begin{equation}
  \psi_{\alpha}    = \frac{\alpha}{4-\alpha} \left ( \frac{r_{H0}}{r_{c}} \right )^{2-\alpha}   , 
\label{eq:psi_a}      
\end{equation}
and $\alpha$ and $\psi_{\alpha}$ are considered to be dimensionless constant parameters.
The crossover scale $r_c$ is likely identified with $r_{H0}$ \cite{Radicella2010}.
When $\alpha =0$, $S_{pl}$ reduces to $S_{\rm{BH}}$. 
In this study, $\psi_{\alpha}$ is assumed to be positive for an accelerating universe \cite{Koma11} and, therefore, $0 < \alpha <4$ is obtained from Eq.\ (\ref{eq:psi_a}).
We note that $\alpha$ and $\psi_{\alpha}$ may be independent free parameters \cite{Koma14}.

Differentiating Eq.\ (\ref{eq:Spl}) with regard to $t$ and applying Eqs.\ (\ref{eq:SBH2}) and (\ref{eq:dSBH}) gives  \cite{Koma11}
\begin{align}
\dot{S}_{pl}  
                 &=  \dot{S}_{\rm{BH}}  \left [  1-   \frac{\psi_{\alpha} H_{0}^{2- \alpha} }{H^{2- \alpha} } \right ]  + {S}_{\rm{BH}}  \left [ \frac{ (2-\alpha) \psi_{\alpha} H_{0}^{2-\alpha} \dot{H} }{H^{3-\alpha}}  \right ] \notag \\
                 &=  \frac{-2K \dot{H} }{H^{3}} \left [  1-   \frac{\psi_{\alpha} H_{0}^{2- \alpha} }{H^{2- \alpha} }  \right ]  +  \frac{K}{H^{2}}  \frac{ 2(1-\frac{\alpha}{2}) \psi_{\alpha} H_{0}^{2-\alpha} \dot{H} }{H^{3-\alpha}}   \notag \\
                 &=  \frac{-2K \dot{H} }{H^{3}} \left [ 1-  \frac{ (2- \frac{\alpha}{2}) \psi_{\alpha} H_{0}^{2-\alpha}}{H^{2-\alpha}}  \right ]    \notag \\
                 &=  \dot{S}_{\rm{BH}}  \left [ 1-  \left ( \frac{4- \alpha}{2}  \right ) \psi_{\alpha}  \left ( \frac{H_{0}}{H} \right )^{2-\alpha}  \right ]     .
\label{eq:dSpl}      
\end{align}

We now apply $S_{pl}$ to the present model and calculate two extra driving terms, $f_{\Lambda,pl}(t)$ and $h_{\textrm{B},pl}(t)$.
For this, we first calculate $ ( \partial S_{\Delta} / \partial S_{\rm{BH}} ) $.
Using Eq.\ (\ref{eq:dSHdSBH_1-dS}), replacing $S_{H}$ by $S_{pl}$, and applying Eq.\ (\ref{eq:dSpl}) yields
\begin{align}
                    \left ( \frac{\partial S_{\Delta} }{\partial S_{\rm{BH}}}  \right ) &=\left ( \frac{\partial S_{H} }{\partial S_{\rm{BH}}}  \right ) -1  =  \frac{\dot{S}_{H}  }{\dot{S}_{\rm{BH}}}  -1  = \frac{\dot{S}_{pl}  }{\dot{S}_{\rm{BH}}} -1       \notag \\
                                                                                           &=   -  \left ( \frac{(4- \alpha)\psi_{\alpha}}{2}  \right )  \left ( \frac{H_{0}}{H} \right )^{2-\alpha}      \notag \\
                                                                                           &=   -  \left ( \frac{ \alpha  \Psi_{\alpha}  }{2}  \right ) \left ( \frac{H}{H_{0}} \right )^{\alpha-2}   .  
\label{eq:dSpl_dSBH}      
\end{align}
Here $(4- \alpha) \psi_{\alpha}$ has been replaced by $ \alpha \Psi_{\alpha}$, using a dimensionless positive constant $\Psi_{\alpha}$, to obtain a simple formulation equivalent to a power-law term examined in previous work \cite{Koma11,Koma12,Koma14}.

Integrating Eq.\ (\ref{eq:dSpl_dSBH}) with regard to $H^2$ and applying $d (H^2) = 2H dH$ yields
\begin{align}
              \int \left ( \frac{\partial S_{\Delta} }{\partial S_{\rm{BH}}}  \right ) d (H^2)  
                                                                                                         &=   - \int   \left ( \frac{\alpha  \Psi_{\alpha} }{2}  \right )   \left ( \frac{H}{H_{0}} \right )^{\alpha-2}     2H dH     \notag \\
                                                                                                         &=   -  \Psi_{\alpha} H_{0}^{2} \left ( \frac{H}{H_{0}} \right )^{\alpha}       +C_{0}   ,
\label{eq:Int_dSpl_dSBH}      
\end{align}
where $C_{0}$ is an integral constant.
Substituting Eq.\ (\ref{eq:Int_dSpl_dSBH}) into Eq.\ (\ref{eq:fL(t)_1stLaw_2}) yields 
\begin{equation}
        f_{\Lambda,pl} (t) =  C_{1} + \Psi_{\alpha} H_{0}^{2} \left (  \frac{H}{H_{0}} \right )^{\alpha}  ,
\label{eq:fL_Spl}     
\end{equation}
where $C_{1}$ is $C - C_{0}$ and is considered to be non-negative.
The second term on the right side is a power-law term proportional to $H^{\alpha}$.
Similarly, using $S_{pl}$, we calculate $h_{\textrm{B},pl}(t)$.
Substituting Eq.\ (\ref{eq:dSpl_dSBH}) into Eq.\ (\ref{eq:hB(t)_1stLaw_2}) yields 
\begin{align}
        h_{\textrm{B},pl}(t) &= -    \dot{H} \left ( \frac{\partial S_{\Delta} }{\partial S_{\rm{BH}}}  \right )                    
                                    =     \dot{H} \left (\frac{\alpha \Psi_{\alpha}  }{2} \right )  \left (  \frac{H}{H_{0}} \right )^{\alpha-2}  .
\label{eq:hB_Spl}     
\end{align}
The $h_{\textrm{B},pl}(t)$ term includes $\dot{H}$.
Equations\ (\ref{eq:fL_Spl}) and (\ref{eq:hB_Spl}) imply that $f_{\Lambda,pl} (t)$ is non-negative, whereas $h_{\textrm{B},pl}(t)$ is non-positive when $H > 0$ and $\dot{H} < 0$ are considered from observations \cite{Hubble2017}, where $\Psi_{\alpha} >0$ and $0 < \alpha <4$ are also used.

Finally, we discuss the ratio of the two terms.
Dividing Eq.\ (\ref{eq:hB_Spl}) by Eq.\ (\ref{eq:fL_Spl}) and setting $C_{1}=0$ for simplicity yields 
\begin{align}
      \frac{  h_{\textrm{B},pl}(t)}{  f_{\Lambda,pl} (t) } &=  \frac{\dot{H} \left (\frac{\alpha \Psi_{\alpha}  }{2} \right )  \left (  \frac{H}{H_{0}} \right )^{\alpha-2}  }{ C_{1} + \Psi_{\alpha} H_{0}^{2} \left (  \frac{H}{H_{0}} \right )^{\alpha}  } = \frac{\alpha}{2} \frac{\dot{H}}{H^2}  .
\label{eq:hB_fl_Spl}     
\end{align}
The solution to Eq.\ (\ref{eq:hB_fl_Spl}) is negative when $\dot{H} < 0$. 
In this case, applying $\dot{H} \ge -2 H^{2}$ to Eq.\ (\ref{eq:hB_fl_Spl}) gives
\begin{align}
   - \alpha \leq   \frac{  h_{\textrm{B},pl}(t)}{  f_{\Lambda,pl} (t) } \leq 0 .
\label{eq:hB_fl_Spl_ineq}     
\end{align}
Here $\dot{H} \ge -2 H^{2}$ is assumed for a non-negative horizon temperature, as examined in Sec.\ \ref{Entropy and temperature}.
Equation\ (\ref{eq:hB_fl_Spl_ineq}) indicates that $|f_{\Lambda,pl} (t)|$ is larger than $|h_{\textrm{B},pl}(t)|$ when $0< \alpha <1$.
In particular, the $f_{\Lambda,pl} (t)$ term tends to be dominant when a small positive $\alpha$ is considered.

In this way, we can study the two terms in the present model for a power-law-corrected entropy.
Of course, the thermodynamic constraints on the two terms can be examined by applying results in Sec.\ \ref{Thermodynamic constraints}.
In addition, we can discuss the background evolutions of the universe in this specific model from a thermodynamic viewpoint.
These tasks are left for future research.

\end{document}